\documentclass[fleqn,usenatbib]{mnras}

\usepackage{newtxtext,newtxmath}
\usepackage[T1]{fontenc}
\usepackage{ae,aecompl}
\usepackage{graphicx}	
\usepackage{amsmath}	
\usepackage{amssymb}	

\title[Weakly relativistic pair shock]{Electrostatic and magnetic instabilities in the transition layer of a collisionless weakly relativistic pair shock}

\author[M. E. Dieckmann et al.]{
M. E. Dieckmann$^{1}$\thanks{E-mail: mark.e.dieckmann@liu.se}
A. Bret$^{2,3}$
\\
$^{1}$Department of Science and Technology (ITN), Link\"oping University, 60174 Norrk\"oping, Sweden.\\
$^{2}$ETSI Industriales, Universidad de Castilla-La Mancha, 13071 Ciudad Real, Spain.\\
$^{3}$Instituto de Investigaciones Energ\'eticas y Aplicaciones Industriales, Campus Universitario de Ciudad Real, 13071 Ciudad Real, Spain.
}
\date{}

\begin{document}
\label{firstpage}
\pagerange{\pageref{firstpage}--\pageref{lastpage}}
\maketitle

\begin{abstract}
Energetic electromagnetic emissions by astrophysical jets like those that are launched during the collapse of a massive star and trigger gamma-ray bursts (GRBs) are partially attributed to relativistic internal shocks. The shocks are mediated in the collisionless plasma of such jets by the filamentation instability of counterstreaming particle beams. The filamentation instability grows fastest only if the beams move at a relativistic relative speed. We model here with a particle-in-cell (PIC) simulation the collision of two cold pair clouds at the speed c/2 (c: speed of light). We demonstrate that the two-stream instability outgrows the filamentation instability for this speed and is thus responsible for the shock formation. The incomplete thermalization of the upstream plasma by its quasi-electrostatic waves allows other instabilities to grow. A shock transition layer forms, in which a filamentation instability modulates the plasma far upstream of the shock. The inflowing upstream plasma is progressively heated by a two-stream instability closer to the shock and compressed to the expected downstream density by the Weibel instability. The strong magnetic field due to the latter is confined to a layer 10 electron skin depths wide. 
\end{abstract}


\section{Introduction}

Binaries with an accreting compact object, which are known as microquasars, can emit relativistic jets \citep{Falcke96,Mirabel99,Madejski16}. The jets are composed of electrons, positrons \citep{Siegert16} and an unknown fraction of ions and they are sources of intense electromagnetic radiation, which can be observed over astronomical distances. Some of this radiation is attributed to synchrotron emissions by relativistic electrons and positrons that gyrate in a strong magnetic field. The source of the jet's magnetic field is under debate. A plasma is a good conductor and, hence, the jet will carry with it the strong magnetic field that is present at its base. Instabilities, which develop in a non-equilibrium collision-less plasma, can also magnetize the jet.

Instabilities can be driven by a large variation of the jet's mean velocity, which is caused by a non-uniform plasma acceleration efficiency of its engine. A velocity variation steepens if a faster plasma cloud catches up with a slower one, which ultimately results in the formation of an internal shock in the jet. Shocks in the jet heat up its plasma and compress the magnetic field. We expect that these internal shocks are powerful sources of electromagnetic emissions. The internal shock model has initially been invoked by \citet{Rees78} to explain the emissions by the knots in the jets of active galactic nuclei and the internal shock model also explains the prompt emissions of gamma-ray bursts \citep{Piran99,Piran04}. \citet{Malzac14} and \citet{Drappeau15} have examined to what extent the emissions of the relativistic jets of microquasars can be explained by internal shocks.

The large mean free path of the plasma particles in the jet implies that Coulomb collisions between particles can not rapidly establish a stable Maxwellian velocity distribution at every point. An anisotropic particle velocity distribution can survive for a long time and drive velocity-space instabilities, which let electromagnetic fields grow at the expense of the thermal anisotropy. The interaction of the particles with these fields speeds up the equilibration of the velocity distribution. The waves, which are driven by the instabilities, are often the ones that sustain a shock in collisionless astrophysical plasma. \citet{Marcowith16} provide a recent review of such shocks. Wave instabilities can, however, also grow ahead of a shock due to downstream particles that leak upstream or they can be driven behind a shock by a plasma that has not been fully thermalized by the shock crossing.

Many prior studies have considered instabilities, which are driven by electrons or by electrons and positrons, that are strong enough to trigger the formation of a shock. \citet{Fried59} demonstrated that magnetowaves grow if two beams of charged particles move at a high relative speed. In what follows, we call this instability the filamentation instability. The filamentation instability has been invoked as a means to magnetize colliding pair clouds \citep{Medvedev99,Brainerd00}. It has been examined for colliding or interpenetrating pair clouds numerically by \citet{Kazimura98,Silva03,Jaroschek04,Milos06,Dieckmann06,Dieckmann09b,Sironi09,Sironi13} and experimentally by \citet{Tatarakis03,Huntington15}. It is now possible to produce relativistically moving clouds of pair plasma \citep{Sarri13a} and it will eventually become possible to study in the laboratory the instabilities that develop between pair beams and the background electrons. These studies show that the filamentation instability drives the strong magnetic fields that mediate the relativistic pair shocks. 

The Weibel instability \citep{Weibel59} affects an electron plasma with a temperature that is lower in one direction than in the plane orthogonal to it. \citet{Morse71} confirmed its existence by means of a particle-in-cell (PIC) simulation. The Weibel instability has been examined more recently by PIC simulations in unbounded \citep{Okabe03,Stockem09,Kaang09} and bounded plasma \citep{Schoeffler14} and it has been observed experimentally by \citet{Quinn12}. The electron Weibel instability is important because it grows already for small thermal anisotropies and, thus, in an almost thermal plasma. It has, for example, been invoked as a means to magnetize the galactic plasma \citep{Ryu12}. 

Most previous PIC simulation studies have created pair shocks by letting equally dense pair clouds collide at a highly relativistic speed. These shocks are mediated by the filamentation instability because its exponential growth rate exceeds in this case that of all other instabilities. However, not all collisions within pair jets will necessarily involve relativistic speeds. This is true in particular for the jets of microquasars that expand only at a moderately relativistic speed \citep{Falcke96}. If the relative speed between the beams is not highly relativistic then the magneto-instabilities compete with the predominantly electrostatic two-stream modes (See \citet{Bret10} for a review). 

\citet{Dieckmann17} studied a shock that emerged out of the collision of two pair clouds at the speed 0.2c (c: speed of light). The two-stream instability grew fastest and it mediated the shock. The quasi-electrostatic two-stream modes in the shock transition layer could, however, not fully thermalize the inflowing upstream plasma. The particle temperature behind the shock was higher along the shock normal than orthogonal to it, which triggered the Weibel instability. Magnetic field patches formed behind the shock that would ultimately establish a thermal equilibrium in the downstream plasma. The magnetic fields were, however, too weak and too far behind the shock to influence it. 

The filamentation instability becomes more important compared to the two-stream instability when we increase the collision speed. There has to be a regime, in which electrostatic and electromagnetic processes attain a similar importance for the shock evolution. We examine here the shock that emerges out of the collision of two equal pair clouds at the speed 0.5c. The exponential growth rate of electrostatic instabilities exceeds that of the filamentation instability for a collision speed $\le$ 0.57c \citep{Bret16} and the shock forms on the electrostatic time scale. The interaction of the two-stream modes with the pair plasma mixes the particles of the counterstreaming beams, which inhibits the growth of the filamentation instability and compresses the plasma density to a value above that expected from a beam superposition. 

The phase space mixing of the particles takes place mainly along the beam direction and the resulting particle population is anisotropic in velocity space. The Weibel instability starts to grow and the magnetic field energy reaches a near-equipartition with the thermal energy in a thin layer. This magnetic field thermalizes the pair plasma and it compresses it to the value expected for a strong two-dimensional shock. The shock structure becomes similar to that observed in previous simulations of relativistic pair shocks. 

Some of the energetic particles, which are generated by the two-stream instability ahead of the shock, escape upstream. This hot particle beam interacts with the incoming upstream plasma. A filamentation instability grows far ahead of the shock, which creates current filaments in the electron and positron distributions. The spatial distributions and the velocity distributions of both species match and the current contributions of both species cancel out each other. This filamentation does thus not result in a strong net current and, hence, in no significant magnetic field. The inflowing upstream plasma is heated by this instability. The growth rate of the two-stream instability decreases if the interacting beams are hot, which explains why it becomes less important at late times.

Our paper is structured as follows. Section 2 discusses the relevant instabilities, it provides an analytic estimate for the shock formation time and the particle-in-cell (PIC) simulation method and the initial conditions are presented. Section 3 presents the simulation results and our findings are summarized in Section 4.

\section{Theory and simulation setup}

\subsection{Initial conditions and shock formation time}

\begin{figure}
\includegraphics[width=\columnwidth]{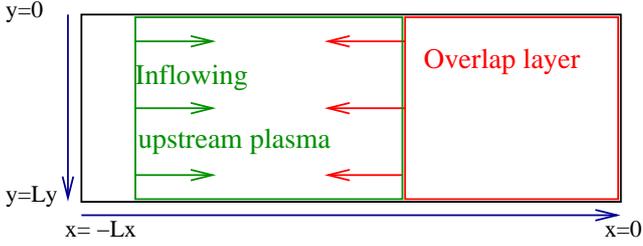}
\caption{The simulation setup: a pair plasma collides with a reflecting wall at $x=0$. The pair cloud has the initial length $L_x$ along $x$ and it moves at the speed $v_0$ to the right. A layer grows close to the wall in which the inflowing and reflected plasma particles coexist. The front of the overlap layer expands at the speed $-v_0$ to decreasing values of $x$ until instabilities set in.}\label{Setup}
\end{figure}

We investigate the collision of a pair plasma with a reflecting wall located at the position $x=0$ at a normal incidence in a two-dimensional (2D) geometry, neglecting Coulomb collisions between particles. The plasma cloud initially fills the entire simulation box, which has the length $L_x$ along $x$ and the length $L_y$ along $y$. The plasma consists of spatially uniform comoving electrons and positrons with the same number density $n_0$ and temperature $T_0=1.16 \times 10^5$ Kelvin. The plasma moves at the mean speed $v_0=c/4$ towards increasing x. The plasma is thus charge-neutral and current-neutral and initially the electric field and magnetic field are zero everywhere in the simulation box. Both species have a Maxwellian velocity distribution with the thermal speed $v_{th}={(k_BT_0/m_e)}^{1/2}$ or $v_{th}/c = 4.4 \times 10^{-3}$. The electron mass is $m_e$ and $k_B$ is the Boltzmann constant. The reflected plasma and the inflowing upstream plasma interpenetrate at the relative speed $c/2$ and form an overlap layer close to $x=0$ that expands at the speed $-v_0$ to decreasing values of $x$ as shown in Fig. \ref{Setup}. The electron density and positron density in the overlap layer are $2n_0$ each. 

\begin{figure}
\includegraphics[width=\columnwidth]{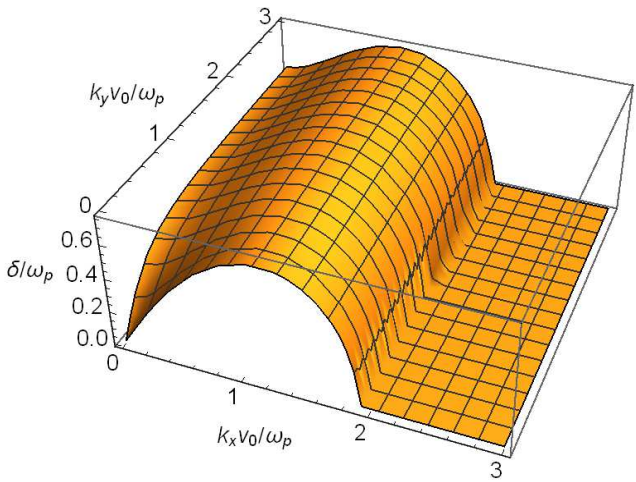}
\caption{Solution of the linear dispersion relation: the growth rate of the unstable modes is plotted as a function of $k_x v_0/\omega_p$ and $k_y v_0/\omega_p$. The electrostatic two-stream instability with $k_x v_0 / \omega_p \approx 1$ governs the unstable spectrum.}\label{gr}
\end{figure}

The plasma in the overlap layer is unstable against the two-stream and filamentation instabilities, among others. The stability of our unmagnetized system is equivalent to the one of two counter-streaming and nearly cold ($v_0/v_{th}=56$) electron beams. The growth rate $\delta$ of any mode can be computed in terms of its wave vector $\mathbf{k}$ following a standard procedure \citep{Bret04,Bret10}. For the present parameters, the result is displayed in Fig. \ref{gr} in terms of the reduced wave-vector components $k_x v_0/\omega_p$ and $k_y v_0/\omega_p$ with
\begin{equation}\label{eq:wp}
\omega_p = {(n_0e^2/m_e\epsilon_0)}^{1/2}.
\end{equation}
As evidenced by Fig. \ref{gr}, the system is governed by the two-stream instability. Its maximum growth-rate is given by
\begin{equation}\label{eq:GR_TS}
  \delta = \frac{\omega_p}{\sqrt{2}} = 0.7 \omega_p.
\end{equation}
In what follows we call the two-stream mode with $k_x v_0/\omega_p \approx 1$ and $k_y=0$ the electrostatic two-stream mode and the modes with $k_x v_0/\omega_p \approx 1$ and $k_y\neq 0$ the oblique modes. The oblique modes are quasi-electrostatic in the sense that the waves interact with the particles primarily via the electric fields. The term two-stream mode encompasses the electrostatic two-stream mode and the oblique modes. 

In contrast, the maximum growth rate of the quasi-magnetic Weibel instability found for $k_x=0$ reads,
\begin{equation}\label{eq:GR_Wei}
  \delta =  2 \omega_p \frac{v_0}{c} = 0.5 \omega_p.
\end{equation}

The inflowing and reflected plasma get mixed once the unstable waves reach a large-enough amplitude and a shock forms close to the front of the overlap layer that moves to the left in Fig. \ref{Setup}. The time it takes for the shock to form can be estimated following the guidelines already laid out for the relativistic regime \citep{Bret13,Bret14}. The dominant instability grows and saturates at the time $\tau_s$. The electromagnetic fields in the overlap layer block the inflowing upstream plasma after their saturation and the shock starts to form close to the boundary between the overlap layer and the inflowing upstream plasma. The overlap layer will gradually change into the downstream region.

A fully developed shock in a plasma with two degrees of freedom sustains a density jump of 3 between the upstream and downstream plasmas \citep{Zeldovich67}. Additional plasma thus has to flow into the overlap layer in order to increase the density of the electrons or positrons from $2n_0$ to $3n_0$. If it took $\tau_s$ to bring the central region density from $n_0$ to $2n_0$, and if the size of this region no longer increases because the flow is blocked within, then one has to wait another $\tau_s$ for the central density to go from $2n_0$ to $3n_0$. In total, the shock formation time is thus about $2\tau_s$.

Let us now turn to the analytical evaluation of $\tau_s$ in the rest frame of the wall. The speed modulus of each pair cloud is $\beta=v_0/c =1/4$. The two-stream instability has the largest exponential growth rate given by Eq. (\ref{eq:GR_TS}). The flow-aligned component of the wave vector of the fastest-growing two-stream mode $k_{x,m}$ is
\begin{equation}\label{eq:kmax_TS}
  k_{x,m} =\sqrt{\frac{3}{2}}\frac{\omega_p}{v_0}.
\end{equation}

The saturation time can be evaluated stating that the electric field grown by the instability, goes from $E_i$ to $E_s$ within the time $\tau_s$. We thus write,
\begin{equation}\label{eq:tau_s}
  E_s =E_i \exp (\delta \tau_s) ~~\Rightarrow ~~ \tau_s = \delta^{-1} \ln \left( \frac{E_s}{E_i} \right).
\end{equation}
The field at saturation $E_s$ can be derived stating the instability saturates when the rebound frequency in the growing field becomes comparable to the growth rate
\begin{equation}\label{eq:E_s}
  \frac{k_m q E_s}{m} = \delta ^2 ~~\Rightarrow ~~ E_s = \frac{1}{\sqrt{6}}\frac{m v_0 \omega_p}{q}
\end{equation}
as discussed by \citet{Bret10}.

The initial field that triggers the instability comes from the spontaneous fluctuations of the plasma. We can evaluate their amplitude following the reasoning of \citet{Buneman59}. In the plasma's rest frame the thermal energy $k_BT$ is carried by wave modes with $k < 1/\lambda_D$ where $\lambda_D = v_{th}/\omega_p$ is the Debye length. Among all these modes, which therefore occupy in our 2D setup a disk of radius $1/\lambda_D$, only modes with $k_x < \sqrt{3/2}\omega_p/v_0$ are picked-up by the instability mechanism. These modes are located in a stripe of width $2\sqrt{3/2}\omega_p/v_0$ along the $k_x$ direction, and  $2\lambda_D^{-1}$ high along the $k_y$ direction. The fraction of $k_BT$ carried by the modes exciting the instability is therefore,
\begin{equation}\label{eq:eps}
  \epsilon = \frac{2\sqrt{3/2}(\omega_p/v_0) ~  2\lambda_D^{-1}}{\pi (\lambda_D^{-1})^2} = \frac{4}{\pi}\sqrt{\frac{3}{2}}\frac{v_{th}}{v_0}.
\end{equation}
Assuming that the amplitude $E_i$ of the fluctuations satisfies $E_i^2/8\pi = \epsilon n_0 k_BT$, we get
\begin{equation}\label{eq:E_i}
E_i^2 = 32\sqrt{\frac{3}{2}}\frac{v_{th}}{v_0} n_0 k_BT.
\end{equation}
From Eqs. (\ref{eq:GR_TS},\ref{eq:tau_s}), the shock formation time $\tau_f = 2\tau_s$ finally reads,
\begin{equation}\label{eq:TSat}
 \tau_f = \omega_p^{-1} \sqrt{2} \ln \left( \frac{\pi}{24\sqrt{6}} \frac{v_0^3}{v_{th}^3} \right).
\end{equation}

The ratio $v_0/v_{th} = 56$ used in our simulation should result in a formation time $\tau_f = 13\omega_p^{-1}$ measured from the time when the instability starts to grow. 

\subsection{The simulation code}

We study the collision of the pair plasma with the reflecting wall and the subsequent shock formation with the relativistic electromagnetic particle-in-cell (PIC) simulation code \textit{EPOCH} \citep{Arber15}. The code solves the kinetic equations of collisionless plasma. Each species $i$ of a collisionless plasma is described by a phase space density distribution $f_i(\mathbf{x},\mathbf{v},t)$, where $\mathbf{x}$ and the velocity $\mathbf{v}$ are independent coordinates. The charge density of this species is computed via $\rho_i(\mathbf{x},t) = q_i \int f_i(\mathbf{x},\mathbf{v},t) d\mathbf{v}$ where $q_i$ is the particle charge of species $i$. The current density needed by Amp\`ere's law is $\mathbf{J}_i(\mathbf{x},t) = q_i \int \mathbf{v} \, f_i(\mathbf{x},\mathbf{v},t) d\mathbf{v}$.

A PIC code approximates the phase space density distribution $f_i(\mathbf{x},\mathbf{v},t)$ of each plasma species $i$ by an ensemble of computational particles (CPs). The relativistic Lorentz force equation updates the three velocity components $\mathbf{v}_j$ of each CP with the index $j$. For this purpose, the electromagnetic fields have to be interpolated from the grid to the position $\mathbf{x}_j$ of the CP. Once $\mathbf{x}_j$ and $\mathbf{v}_j$ have been updated in time, the plasma current $\mathbf{J}(\mathbf{x},t)$ is computed by summing up the current contributions of all CPs. The electromagnetic fields are updated via Amp\`ere's equation using $\mathbf{J}(\mathbf{x},t)$. This cycle is repeated for as many time steps $\Delta_t$ as necessary.

A PIC simulation code represents the electric field $\mathbf{E}(\mathbf{x},t)$ and the magnetic field $\mathbf{B}(\mathbf{x},t)$ on a numerical grid that discretizes space $\mathbf{x}$ and time $t$. The fields are updated in time using discretized forms of Amp\`ere's law and Faraday's law
\begin{equation}
\frac{\partial \mathbf{E}(\mathbf{x},t)}{\partial t} = \nabla \times \mathbf{B}(\mathbf{x},t) - \mathbf{J}(\mathbf{x},t),
\end{equation}
\begin{equation}
\frac{\partial \mathbf{B}(\mathbf{x},t)}{\partial t} = -\nabla \times \mathbf{E}(\mathbf{x},t).
\end{equation}
Amp\`ere's law and Faraday's law and the particle equations of motion are normalized.  The conversion coefficients from normalized units to SI units are given in Table \ref{Conversion}. \textit{EPOCH} fulfills Gauss's law and $\nabla \cdot \mathbf{B}=0$ to round-off precision. 
\begin{table}
\begin{tabular}{|l|l|l|}
Normalized units & SI units & Conversion factor \\
\hline
Space $\mathbf{x}$ & $c\mathbf{x}/\omega_p$ & $530 \, \mathrm{m}$ \\
Wave number $\mathbf{k}$ & $\omega_p\mathbf{k}/c$ & $2\times 10^{-3} \, \mathrm{m}^{-1}$\\
Time $t$ & $t/\omega_p$ & $1.8 \times 10^{-6} \, \mathrm{s}$\\
Frequency $\omega$ & $\omega\omega_p$ & $5.64 \times 10^6 \, \mathrm{s}^{-1}$\\
Density $n$ & $n_0n$ & $10^{8}\, \mathrm{m}^{-3}$ \\
Electric field $\mathbf{E}$ & $\omega_pm_ec\mathbf{E}/e$ & $960\, \mathrm{Vm^{-1}}$\\
Magnetic field $\mathbf{B}$ & $\omega_pm_e\mathbf{B}/e$ & $3.2\times{10}^{-6}\, T$\\
Velocity $\mathbf{v}$ & $c\mathbf{v}$&$3\times 10^8\, \mathrm{ms^{-1}}$\\
Momentum $\mathbf{p}$ & $m_ec\mathbf{p}$ & $2.73\times 10^{-22}$\\
Current density $\mathbf{J}$ & $en_0c\mathbf{J}$ & $0.48\, \mathrm{C}\mathrm{m}^{-2}\mathrm{s}^{-1}$\\
Charge density $\rho$ & $en_0 \rho$ & $1.602 \times 10^{-8}\mathrm{C}\mathrm{m}^{-3}$\\
\hline
\end{tabular}
\caption{Conversion of quantities from normalized units to SI units. The conversion factor is given for $n_0=10^{8}\mathrm{m}^{-3}$.}\label{Conversion}
\end{table}

We will give our simulation results in normalized units. The electron skin depth $\lambda_S = c\omega_p^{-1}$ is the characteristic spatial scale of the magnetic filamentation modes that mediate the highly relativistic pair shocks. Our lower plasma collision speed implies that electrostatic waves, which can oscillate on spatial scales comparable to the plasma Debye length $\lambda_D/\lambda_S = 6 \times 10^{-3}$, become important. We must resolve this small scale in our simulation, which limits in turn the largest spatio-temporal scale that is accessible to the simulation. 

We use periodic boundary conditions along $y$, a reflecting boundary condition at $x=0$ and an open boundary condition at $x=-L_x$. The electrons and positrons fill up the simulation box at the simulation's start and their distribution is spatially uniform. The pair cloud moves to increasing values of $x$ at the speed $v_0$. No new CPs are introduced while the simulation is running. We perform 3 simulations. 

The one-dimensional simulation 1 resolves the interval $L_x = 405$ by 67500 grid cells. It represents electrons and positrons by $3.4\times 10^7$ CPs each. The absence of a background magnetic field and the one-dimensional geometry restrict the wave spectrum to electrostatic charge density (Langmuir) waves and to ordinary light waves. The latter do not grow in our simulation and the plasma dynamics is thus completely described by the electrostatic field component $E_x$ and by the projected phase space density distributions $f_e(x,v_x)$ and $f_p(x,v_x)$ of the electrons and positrons. This simulation reveals the structure of a shock that is mediated exclusively by electrostatic two-stream modes.

Simulations 2 and 3 resolve two-dimensional domains. Simulation 2 models the onset of the instabilites in the overlap layer and the initial stage of the shock formation. It resolves a domain with the length $L_x=48$ and $L_y=1.5$ by a total of $8000 \times 250$ grid cells. Electrons and positrons are modeled by $2 \times 10^8$ CPs each. The large number of CPs per cell provides an accurate phase space representation for the particles that can be easily visualized. Simulation 3 extends the domain size to $L_x \times L_y = 405 \times 9$ and resolves it by 67500 grid cells along x and by 1500 grid cells along y. The electrons and positrons are each resolved by $10^9$ CPs, respectively. We can resolve a larger spatio-temporal at an acceptable computational cost with this lower number of CPs. The two-stream and filamentation instabilities that develop in the overlap layer drive TM waves in the two-dimensional simulations and we will analyse the complex in-plane electric field $E_p = E_x + iE_y$ and the out-of-plane magnetic field $B_z$. The plasma dynamics is captured by the projections of the electron phase space density $f_e(x, y, v_x, v_y)$ and of the positron phase space density $f_p(x, y, v_x, v_y)$.

\section{The simulation results}

The first subsection presents the results of simulation 2. We demonstrate that the instabilities, which thermalize the plasma in the overlap layer, result in the growth of wave structures on spatial scales $\sim \lambda_D \ll \lambda_S$. The second subsection works out effects caused by the multi-dimensional wave turbulence by comparing the results from simulations 1 and 3. The last subsection presents in more detail the long-term evolution of the pair shock in simulation 3 and how the magnetic field affects the evolution of the shock. The electrons and positrons behave similarly in our simulations due to their equal mass and we will thus analyze almost exclusively the electron distributions. 

\subsection{Initial evolution}

\begin{figure}
\includegraphics[width=\columnwidth]{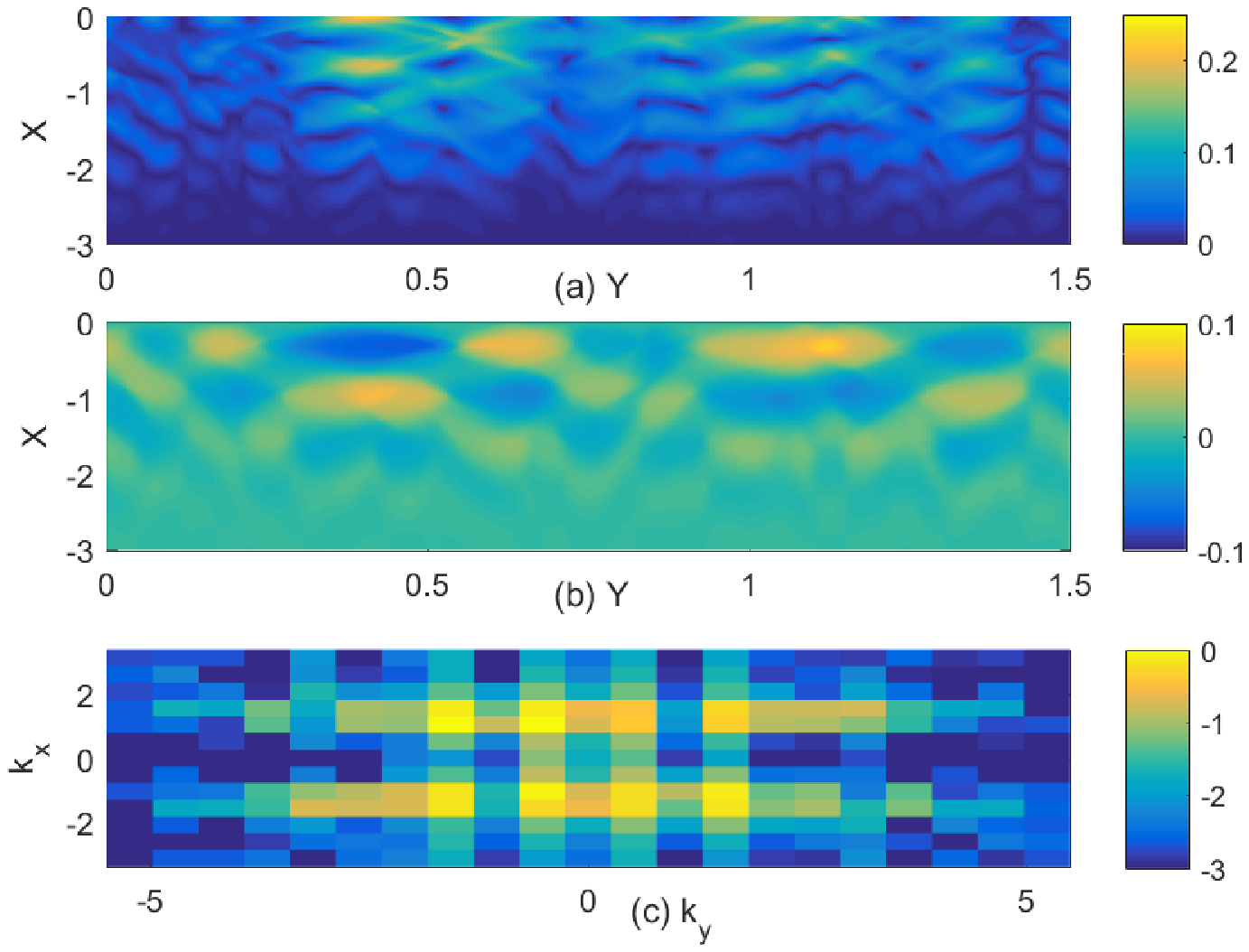}
\includegraphics[width=\columnwidth]{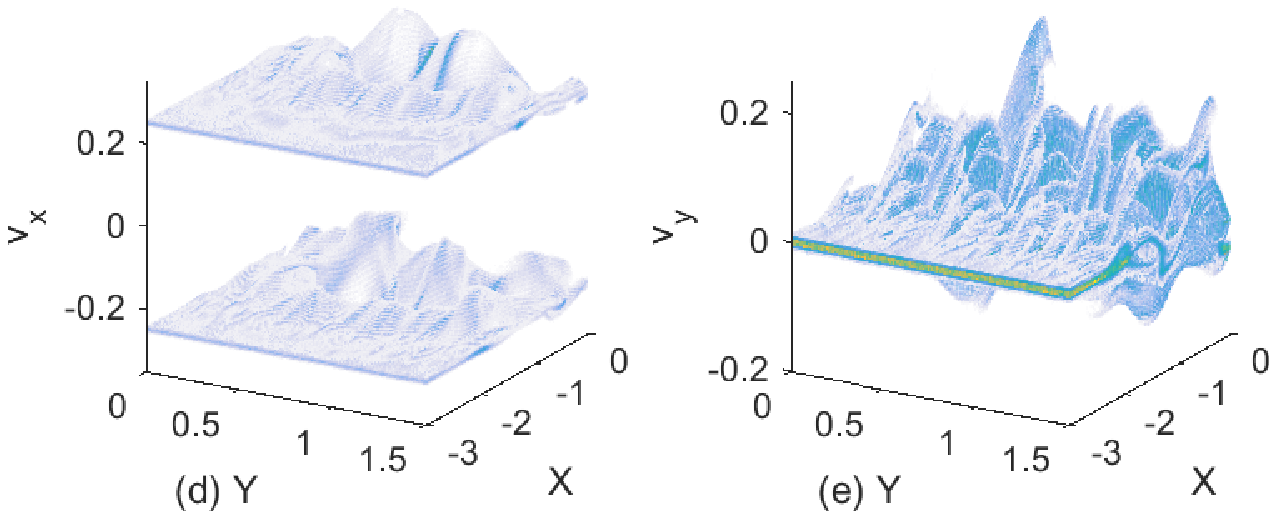}
\caption{The distributions of the electromagnetic fields and of the electrons computed by simulation 2 at the time $t=15$: panel (a) shows the in-plane electric field $|E_p|$. Panel (b) shows the out-of-plane magnetic field $B_z$ and (c) the 10-logarithm of the power spectrum $P(k_x,k_y)$ of $E_p$. The wave numbers are multiplied with $v_0/c$. The projections of the electron phase space density $f_e(x,y,v_x)$ and $f_e(x,y,v_y)$ are shown in (d) and (e), respectively.
}\label{simulation1}
\end{figure}

Figure \ref{simulation1} shows the plasma and field distributions at the time $t=15$. 
The spatial distribution of  $|E_p|$ shows wave activity in the interval $-2.5 \le x \le 0$. These waves are not purely electrostatic because $B_z\neq 0$. However, the electric force that is excerted on a particle exceeds the magnetic force by an order of magnitude, which we can see from the normalization of $\mathbf{E}$ and $\mathbf{B}$ in Table \ref{Conversion} and from the typical particle speed $\sim v_0 = 0.25$. The magnetic and electric fields correlate well in some intervals like in that defined by $0.3 < y <0.7$ and $-1 < x \le 0$, which is typical for oblique modes. 

A comparison of the spatial power spectrum of $E_p$ in Fig. \ref{simulation1}(c) and the solution of the linear dispersion relation in Fig. \ref{gr} confirms that the waves are the oblique modes. Most wave power is concentrated at a wavenumber $|k_x| (v_0/c)\approx 1$ and they have been driven by a resonance between the particles and the waves. The filamentation instability grows in the wave number interval $k_x \approx 0$ and its contribution to the electromagnetic fields is weak at this time. 

Figure \ref{simulation1}(d) shows that the instability has not yet thermalized the electrons close to the wall at $t>\tau_f$, which implies that the instability does not start to grow at $t=0$. The electrons of the inflowing beam with $v_x = 0.25$ are reflected by the wall at $x=0$ and feed the beam with the speed $v_x = -0.25$. The amplitudes of the velocity oscillations along $v_x$ and along $v_y$ are of the order of 0.1 and their wave length is well below $\lambda_S$. These structures are driven by the electric field perturbations that reach wave numbers of up to $k_y (v_0/c) \approx 5$ in Fig. \ref{simulation1}(c), which corresponds to a wave length of $\lambda \approx \pi / 10$ or 20 oscillation periods per $\lambda_S$. 

\begin{figure}
\includegraphics[width=\columnwidth]{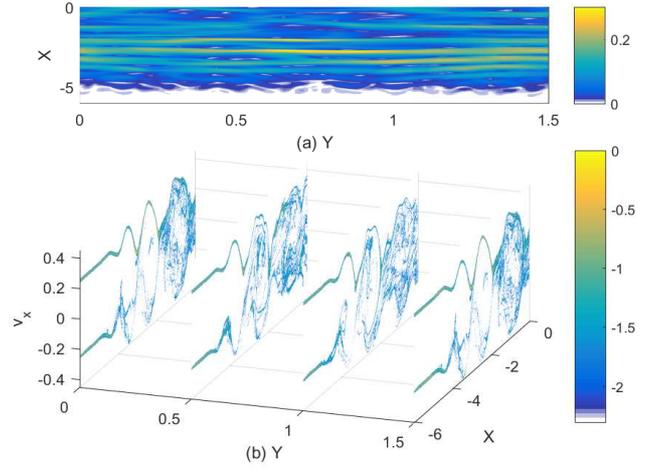}
\caption{The distributions of the electric field and of the electrons computed by simulation 2 at the time $t=25$: Panel (a) shows the in-plane electric field distribution $E_p$ and panel (b) shows slices of $log_{10}(f_e(x,y,v_x))$ for the values $y=0, 0.5,1$ and $y=1.5$.}\label{simulation31}
\end{figure}

Figure \ref{simulation31} shows $|E_p|$ and slices of $f_e(x,y,v_x)$ at the time $t=25$ when the waves have saturated. 
Figure \ref{simulation31}(a) reveals quasi-planar waves with a wave vector that is aligned with the beam velocity vector. Their amplitude is strong enough to force electrons on a circular phase space trajectory in Fig. \ref{simulation31}(b). The structures on the displayed slices are similar and, given the quasi-planarity of the waves, we conclude that they are phase space tubes with an axis that is aligned with $y$. Phase space holes and tubes form when the electrostatic two-stream instability saturates \citep{oneil65,Morse969,eliasson06}. 

\begin{figure}
\includegraphics[width=\columnwidth]{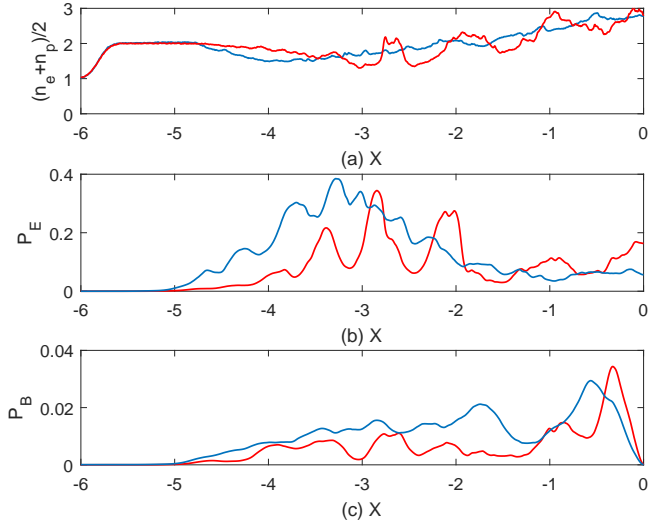}
\caption{A comparison of the field energy densities and particle densities computed by simulations 2 (red curves) and 3 (blue curves) at $t=25$: Panel (a) shows the y-averaged pair density. Panel (b) shows the y-averaged normalized energy density $P_E$ of the in-plane electric field and panel (c) the normalized y-averaged magnetic pressure $P_B$.}\label{extraplot2}
\end{figure}

Figure \ref{extraplot2} compares the y-averaged pair density, magnetic pressure and energy density of the in-plane electric fields computed by the simulations 1 and 2 at the time $t=25$. The pressure $P_B(x) = B_z^2(x)/(2\mu_0P_0)$ of the out-of-plane magnetic field and the energy density of the electric field $P_E = \epsilon_0 (E_x^2+E_y^2)/(2P_0)$ (both in SI units) are expressed in units of the ram pressure $P_0=m_en_0v_0^2$ of the inflowing upstream electrons.  
The energy densities of the magnetic and electric field in simulation 2 are on average below those in simulation 3, which suggests that the instability in simulation 3 has evolved farther than that in simulation 2. Simulation 2 uses 10 times more CPs per cell, which yields statistical noise with a lower amplitude. The instability in simulation 2 requires more time to grow from noise levels to its saturation than its counterpart in the noisier simulation 3. The smaller box of simulation 2 furthermore causes larger fluctuations of the y-averaged $P_E(x)$ and $P_B(x)$.  

\subsection{Shock formation}

Simulation 2 demonstrated that the oblique modes outgrow the magnetic filamentation modes and that they dominate at least initially the plasma dynamics. We can gain additional insight into the importance of magnetic field effects by comparing the plasma evolution in simulation 1, which only resolves the electrostatic two-stream instability, with that in simulation 3 that takes the oblique modes and the magnetic instabilities into account.

\begin{figure}
\includegraphics[width=\columnwidth]{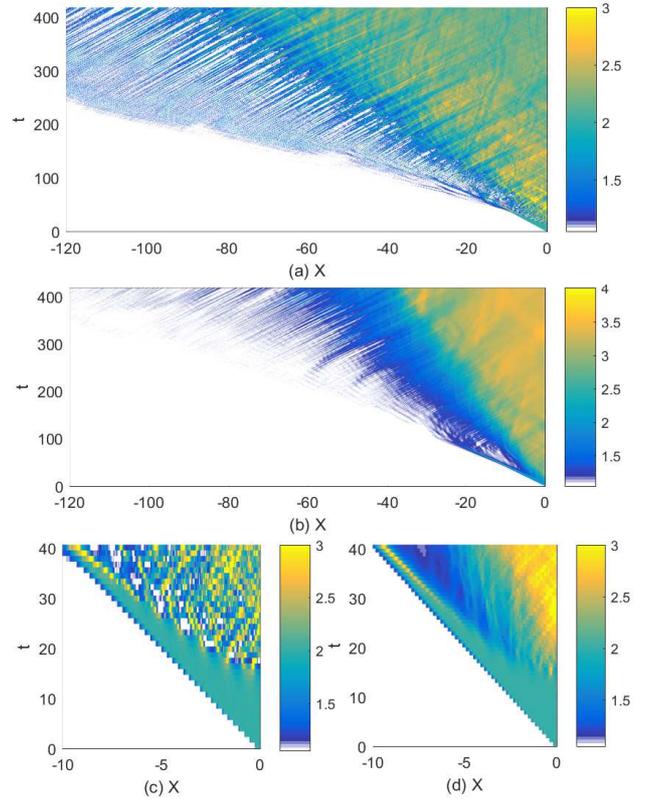}
\caption{The pair densities computed by simulations 1 and 3: panel (a) shows the time evolution of the electron density $n_e(x,t)$ from simulation 1. Panel (b) shows the y-averaged electron density $n_e(x,t)$ from simulation 3. Panel (c) shows the initial density evolution in simulation 1 and panel (d) that in simulation 3. The displayed color range has been chosen such that the inflowing plasma with the mean density $n_0$ and its statistical number density fluctuations is not visible.}\label{DensityTime}
\end{figure}

Figures \ref{DensityTime}(a,b) compare the long-term evolution of the electron density distributions $n_e(x,t)$ computed by simulations 1 and 3. 
A boundary forms in the density distributions of both simulations that separates the inflowing upstream plasma from the plasma in the overlap layer. The density change across the boundary is fast in simulation 1, while the density increases gradually over 10-20 spatial units in simulation 3. The propagation speed of the boundary and the density jump across the boundary are different in both simulations. The boundary in simulation 1 propagates 80 spatial units during 420 time units, giving the speed $\approx$ -0.2. The downstream electron density is 2.2. The propagation speed of the boundary in simulation 3 is -0.1 and the electron density downstream of the shock is about 3.3. 

Figures \ref{DensityTime}(c,d) compare the shock formation in both simulations.  The smoother density evolution in simulation 3 can be explained in parts by the averaging over y because the phase space holes, which are responsible for the density fluctuations, form in both simulations. The boundary, which separates the unperturbed upstream plasma with $n_e=1$ from the overlap layer with $n_e=2$, propagates 10 spatial units during 40 time units and its speed is thus $-v_0$. The density in the overlap layer increases beyond $n_e=2$ after $t\approx 15$, which was the time when the velocity oscillations in Fig. \ref{simulation1} grew to a noticable amplitude. The simultaneous growth of the density beyond $n_e=2$ in both simulations implies that the same instability is at work. Simulation 1 resolves only the electrostatic two-stream mode. This mode grows at the same rate as the oblique modes (See Fig. \ref{gr}) and we thus identify the two-stream modes 
as the ones that trigger the shock formation in simulation 3.
 
The distribution in Fig. \ref{DensityTime}(d) reaches the value $\sim$3, which we expect for the density of the downstream plasma behind a two-dimensional shock, close to the wall at $t\approx 23$. We thus infer with $\tau_f = 13$ that the two-stream instability started to grow at $t=10$. The front of the reflected beam at $-v_0t$ has reached the position $-2.5$ at this time. Figure \ref{simulation1}(d) shows that an x-interval of width $\approx 0.5$ at the front of the reflected beam does not participate in the instability, confining the instability to an interval along x with the thickness $\approx$ 2. Hence the two-stream instability starts to grow once the thickness of this interval is comparable to the wave length $\lambda = \sqrt{3}\approx 1.7$ (Eq.  \ref{eq:kmax_TS}) of the fastest-growing waves. 

\begin{figure}
\includegraphics[width=\columnwidth]{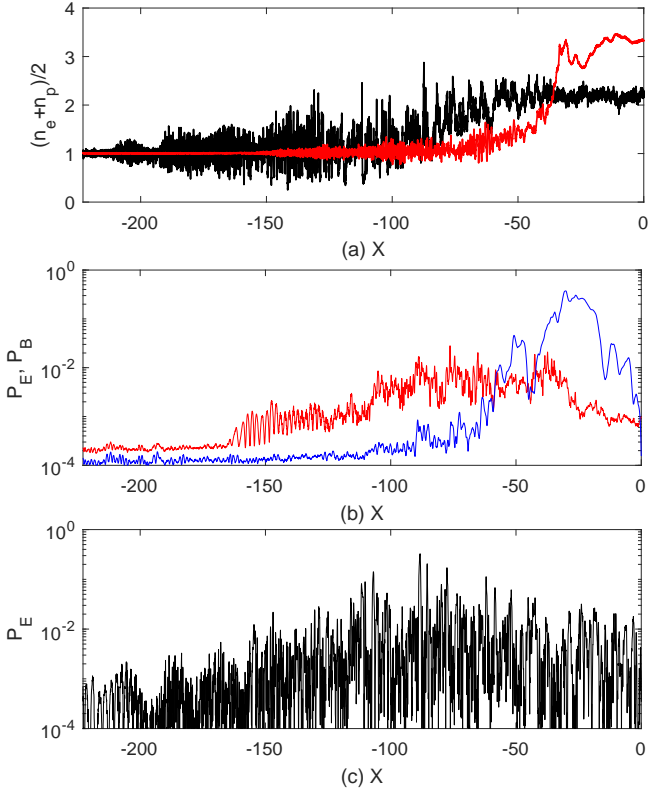}
\caption{A comparison of the simulation results of simulations 1 and 3 at $t=420$: panel (a) shows the y-averaged pair density in the 1D simulation (black curve) with that in the 2D simulation (red curve). Panel (b) shows the y-averaged normalized energy density $P_E$ (red curve) of the in-plane electric field and of $P_B$ (blue curve) in simulation 3. Panel (c) shows $P_E$ computed by simulation 2.}\label{extraplot}
\end{figure}

Figure \ref{extraplot} compares $P_E$, $P_B$ and the pair density in simulations 1 and 3 at the time $t=420$.
The pair density in simulation 1 reaches about 2.2 times its upstream value close to the wall at $x=0$, while that in simulation 3 increases to over 3 times this value. The transition layer is also much broader in simulation 1 and density oscillations have propagated to much lower values of $x$ than in simulation 3. Figure \ref{extraplot}(b) shows that $P_B$ has grown to a much larger value than in Fig. \ref{extraplot2} and that its peak value exceeds now the typical values of $P_E$. The energy density of the electric field still exceeds by far that of the magnetic field in the interval $-170 \le x \le -60$. 

\begin{figure}
\includegraphics[width=\columnwidth]{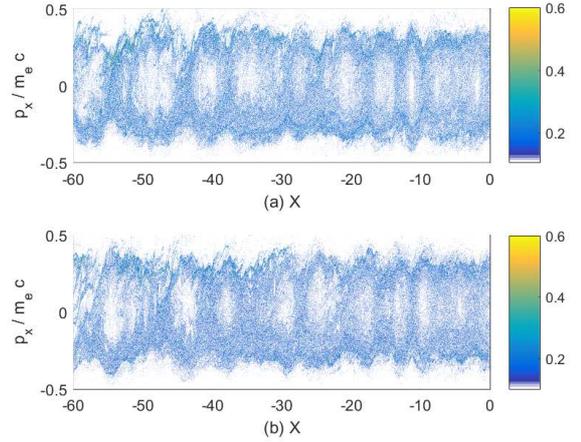}
\caption{The phase space density distributions of the particles close to the reflecting wall in simulation 1 at the time $t=420$: panel (a) shows that of the positrons and panel (b) that of the electrons. Both distributions are normalized to the peak density in the displayed intervals. The linear color scale is clamped to values between 0.1 and 0.6.}\label{extraplot3}
\end{figure}

Confining the simulation geometry to one spatial dimension suppresses the Weibel- and filamentation instabilities because they are driven by the separation of currents in the direction that is orthogonal to the beam direction; only $P_E$ thus grows in simulation 1. Figure \ref{extraplot}(c) shows a broad and strong distribution of $P_E$, which reaches peak values that are comparable to those of $P_E$ and $P_B$ in simulation 3. The small-scale oscillations are tied to electron phase space holes, which are shown by Fig. \ref{extraplot3}.    
These structures are the same as those shown in Fig. \ref{simulation31}. An electron phase space hole is sustained by a positive net charge in its center around which the electrons gyrate. A positron phase space hole is sustained by a negative net charge. The electron- and positron phase space holes are shifted along space in Fig. \ref{extraplot3}. The phase space holes become increasingly diffuse with increasing values of $x$. Periodic trains of phase space holes in one spatial dimension are prone to coalescence instabilities \citep{Roberts} and modulational instabilitities \citep{Schamel1}, which ultimately result in a quasi-thermal plasma \citep{Schamel2}. Quasi-thermal means here that the phase space density distribution of the leptons does no longer show structures like those in Fig. \ref{extraplot3} but that its velocity distribution is not necessarily a Maxwellian as it is the case for a thermal plasma. We would obtain a quasi-thermal plasma at the wall and thus a shock if our simulation would run a few times longer.

The faster expansion and the lower plasma compression in simulation 1 are a consequence of the constraint of the wave and particle dynamics to one spatial dimension. A shock converts the directed flow energy of the upstream plasma into heat. The inflowing upstream plasma carries the same energy density in both simulations. The electrostatic two-stream mode heats the plasma in simulation 1 along x while the ensemble of the two-stream modes in simulation 3 heats up the particles along both resolved directions. 

\begin{figure}
\includegraphics[width=\columnwidth]{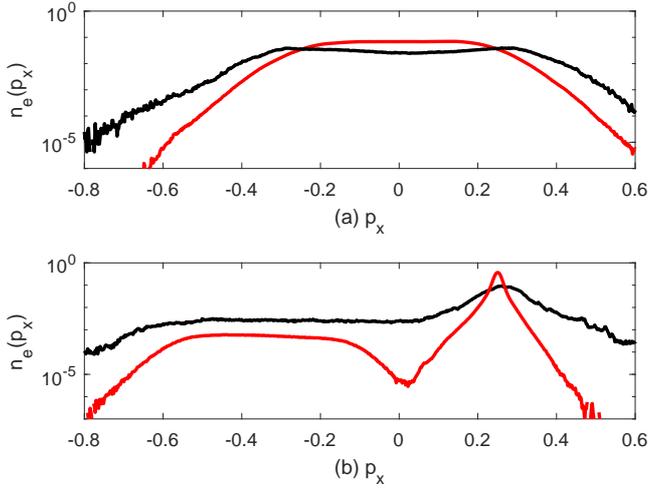}
\caption{The electron velocity distributions computed by simulations 1 and 3 at the time $t=420$: panel (a) compares the electron distributions as a function of the momentum $p_x$ of simulation 1 (black curves) with that in simulation 3 (red curves) averaged over $-15\le x\le0$. Panel (b) compares the velocity distributions averaged over $-120\le x \le -100$. The distributions are normalized to the peak value of the distribution at $t=0$.}\label{simulation1D2Dvel} 
\end{figure}

Figure \ref{simulation1D2Dvel} demonstrates the consequence of the additional degree of freedom for the wave and particle dynamics by comparing the momentum distributions $n_e(p_x)$, which were computed by simulations 1 and 3 at the time $t=420$, in the far upstream and in the downstream. 
The electron density distribution $n_e(p_x)$ in simulation 3 has its maximum at $p_x=0$ and it decreases monotonically with increasing values of $|p_x|$. The equivalent distribution in simulation 1 peaks at values $p_x \approx \pm 0.25$ close to the initial momenta of the incoming and reflected beams. The proportion of electrons with a large value of $|p_x|$ is thus larger in simulation 1 than in simulation 3; the downstream plasma and the shock, which separates it from the upstream plasma, will expand faster in simulation 1. The lower expansion speed of the shock in simulation 3 implies in turn a larger plasma compression.  

Figure \ref{DensityTime} revealed density structures, which crossed the displayed spatial interval during 250 time units (simulation 1) and 400 time units (simulation 3) and outran the shock. Figure \ref{simulation1D2Dvel}(b) shows the distributions $n_e(p_x)$ computed by both simulations in the interval $-120 \le x \le -100$ at $t=420$. We observe in both simulations particles that move upstream with $p_x <0$ and are responsible for the density structures. The distribution of simulation 1 has a plateau in the interval $-0.5 \le p_x \le 0$, which goes over into the incoming beam in the interval $0.2 \le p_x \le 0.3$. Such plateaus are typical for the electron velocity distribution after the electrostatic two-stream instability has saturated \citep{oneil65}. The distribution in simulation 3 shows a narrower plateau in the interval $-0.5 \le p_x \le -0.2$, which is separated from the beam of incoming upstream electrons by a pronounced density minimum at $p_x \approx 0$. The beam at $p_x \approx 0.25$ in simulation 3 is cooler than that in simulation 1. The hot beam that interacts with the inflowing upstream particles is denser in simulation 1 than in simulation 3, which results in a stronger instability and, hence, stronger particle heating upstream of the shock. 

\subsection{Long-term evolution in simulation 3}

\begin{figure}
\includegraphics[width=\columnwidth]{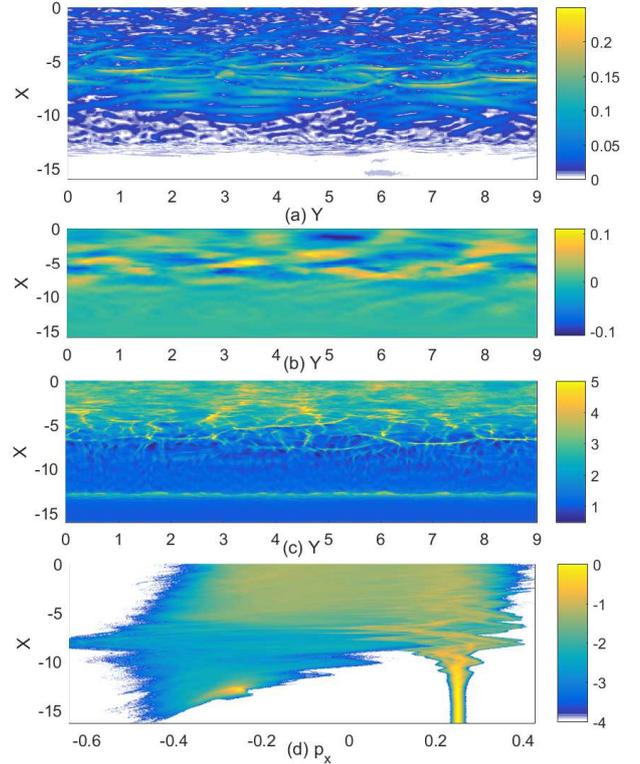}
\caption{The distributions of the electromagnetic fields and of the electrons at the time $t=55$: panel (a) shows the in-plane electric field $|E_p|$. Panel (b) shows the out-of-plane magnetic field $B_z$. The electron density distribution $n_e(x,y)$, which is clamped to a maximum of 5, is displayed in panel (c). The peak density reaches values of 10. Panel (d) shows $f_e(x,p_x)$, which has been averaged over all $y$.}\label{time55} 
\end{figure}

In what follows we examine exclusively the data from simulation 3. Figure \ref{time55} shows the electron distribution and the electromagnetic field distribution close to the shock at the time $t=55$.
We observe quasi-planar waves in the interval $-10\le x \le -4$, which concides with that in Fig. \ref{DensityTime}(b) where the plasma density is gradually increasing. We refer to this interval as the shock transition layer. The electric and magnetic field amplitude of the waves is about 0.1 and the electric field oscillations reach a peak amplitude $\approx 0.25$; the nonrelativistic particles will interact mainly with the electric field. The Larmor radius $v/B_z$ of particles that move with the speed $v\sim v_0$ exceeds the size of the magnetic field patches even for the largest observed value $B_z = 0.1$; the magnetic field deflects the particles but it can not trap them. The low field amplitude for $x>-4$ indicates that the two-stream instability is ineffective in this region.

Figure \ref{time55}(c) shows the density distribution $n_e(x,y)$. The boundary that confines the dense downstream electrons is located at $x\approx -5$, which coincides with the end of the shock transition layer. The downstream region is not yet in a thermal equilibrium, which we can see from the density striations at $(x,y)\approx(-4,3)$ and $(-4,4.5)$. The electron density oscillates also in the shock transition layer and we observe density spikes with $n_e(x,y)>5$, for example at $(x,y)=(-7,7)$. The thickness of these striations is below 0.1. Another planar high-density structure is located at $x\approx -13$, which has the momentum $p_x \approx -0.25$ in Fig. \ref{time55}(d). According to Figs. \ref{simulation1}(d,e) this density stripe corresponds to the front of the electron beam, which has been reflected by the wall and returned upstream without driving an instability. 

Figure \ref{time55}(d) demonstrates that the inflowing electron beam at $p_x \approx 0.25$ and $x\approx -15$ remains intact until $x\approx -10$ when it reaches the shock transition layer. The beam undergoes velocity oscillations, which are caused by the wave electric field. The oscillations along $p_x$ become stronger with increasing values of x until the beam breaks at $x\approx -4$ and is absorbed by the downstream plasma. The breaking of the beam is caused by the collapse of electron phase space holes and probably also by the particle deflection by the magnetic field fluctuations in this interval.

A beam of hot electrons is observed at $p_x<0$ and at low $x$. These particles have a thermal spread $\sim 0.12$ that is comparable to that in the downstream plasma. The thermal spread at the base of the beam corresponds to a temperature of about 10 keV. Their distribution is bounded by an approximately linear profile that starts at $p_x = -0.45$ at $x=-16$ and ends at $p_x = 0$ at $x=-10$. The linear profile suggests that the beam originates from the shock transition layer since faster electrons move farther during a fixed time and those with $p_x=0$ are located at the front of this layer. The velocity spread of the hot electrons is largest at $x=-7$ and they form the vortex that is typical for an electron phase space hole. The phase space densities of the electrons and positrons and the total plasma density are animated in the supplementary movie 1, showing the evolution of the phase space vortices and the pair outflow.

\begin{figure}
\includegraphics[width=\columnwidth]{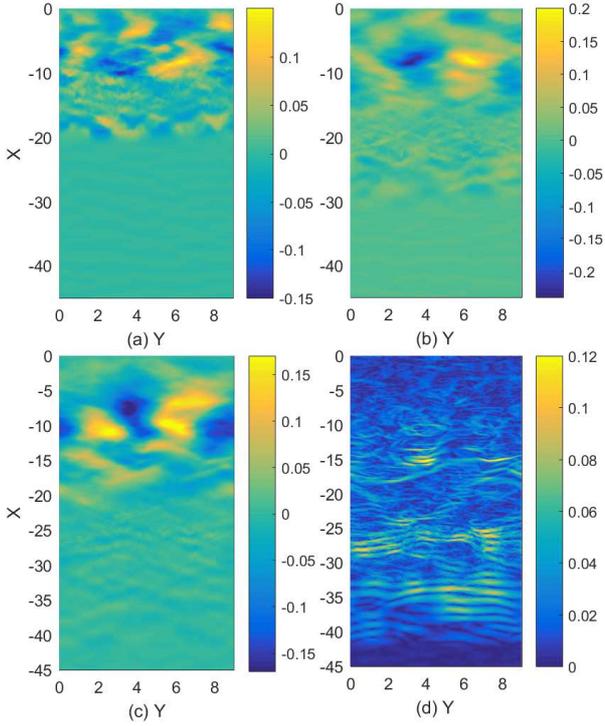}
\caption{Panels (a,b,c) show the out-of-plane magnetic field $B_z$ at the times 80, 110 and 140, respectively. Panel (d) shows the in-plane electric field $|E_p|$ at the time 140.}\label{growth} 
\end{figure}

Figure \ref{growth} shows the distributions of $B_z$ at the times 80, 110 and 140. It also shows $|E_p|$ at the time 140.
The magnetic oscillations have reached at $t=80$ the front of the shock transition layer (See Fig. \ref{DensityTime}(b)). Their amplitude and wavelength in the interval $-18\le x \le -10$ are comparable to those at $t=55$. A magnetic structure is emerging at $(x,y)\approx (-9,5)$. Its amplitude exceeds those in the remaining simulation box by about 50\% and its wavelength along y at $x\approx-9$ is comparable to $L_y$ and therefore much larger than the magnetic field structures driven by the two-stream modes. 
The large magnetic structure continues to grow and reaches the amplitude 0.2 at the time 110. The magnetic structure expands along y and we observe a second oscillation along y at $x\approx -10$ in Fig. \ref{growth}(c). The magnetic structure in Fig. \ref{growth}(c) has no counterpart in $E_p$ (See Fig. \ref{growth}(d)) and it is thus purely magnetic; another instability is at work. 

The inflowing upstream electrons (and positrons) are heated by their interaction with the electric field. Its quasi-planar profile and polarization along x results in anisotropic heating that leads to a higher thermal energy along x than along y. \citet{Weibel59} predicts that in this case the wave vector of the most unstable waves is aligned with the direction y, which is observed in Figs. \ref{growth}(a-c). At $t=140$ the magnetic field of the Weibel mode is much stronger than that of the two-stream modes in the interval $-40 \le x \le -25$ and the latter are accompanied by an $|E_p|>0$.      

\begin{figure}
\includegraphics[width=\columnwidth]{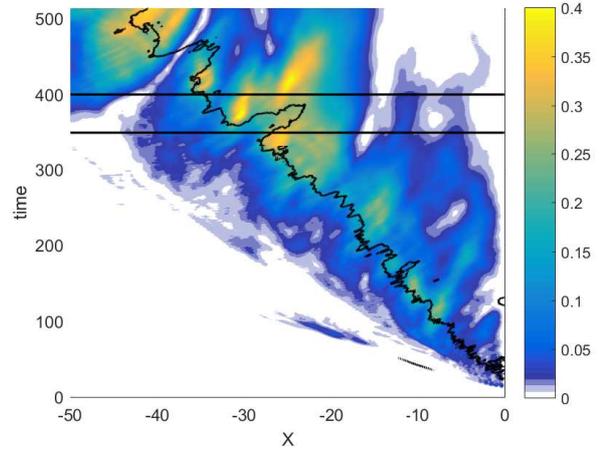}
\caption{The y-averaged and normalized magnetic pressure $P_B (x,t)$. The overplotted curve corresponds to the contour line $n_e(x)=2.8$, where the density has been y-averaged. The horizontal lines denote the times $t=$ 350 and 400.}\label{magpressure} 
\end{figure}

We compare in Fig. \ref{magpressure} the time-evolution of the y-averaged magnetic pressure $P_B(x)$ to the shock front. We define the latter via the contour $n_e(x,y)=2.8$.  
The Weibel modes grow just behind the shock front. The magnetic pressure reaches a peak value of 0.5 and it does not fall below 0.1 after t=100. The thickness of the magnetic layer downstream is about 10 spatial units, which exceeds by far the gyroradii of the fastest particles. The magnetic field is thus strong enough the thermalize the upstream particles that were not scattered by the two-stream waves, which explains why the density reaches its maximum at the magnetic barrier. The magnetic pressure is not negligible in an interval to the left of the shock and this field is tied to the two-stream modes. The magnetic pressure at the shock is not constant in time and we observe in Fig. \ref{magpressure} a sudden drop of $P_B(x,t)$ at $t\approx 350$. The absorption of the magnetic energy by the surrounding plasma heats it. The increased thermal pressure and the subsequent thermal expansion together with the magnetic field redistribution cause an inward motion of the shock front after $t=350$.  

\begin{figure*}
\includegraphics[width=\textwidth]{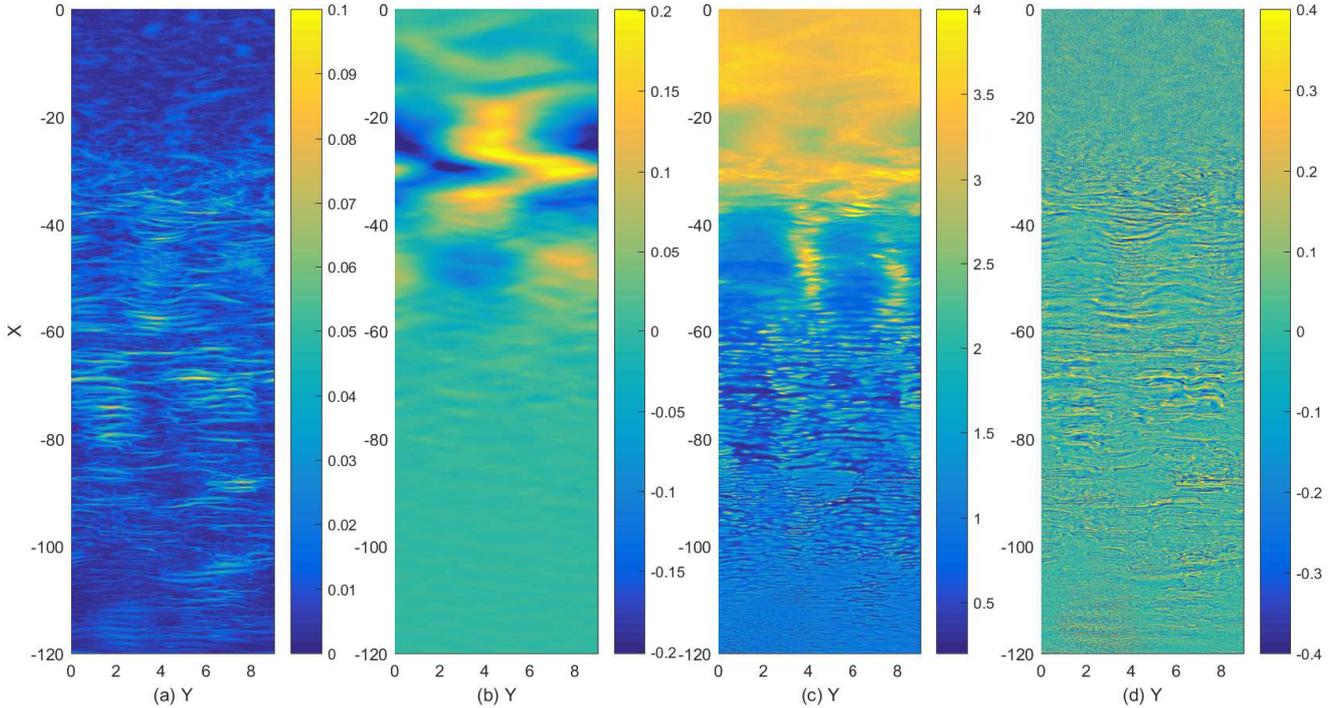}
\caption{The distributions of the electromagnetic fields and of the electrons at the time $t=400$: panel (a) shows the spatial distribution of $|E_p(x,y)|$. That of $B_z(x,y)$ is shown in panel (b). Panel (c) shows the electron density distribution $n_e(x,y)$. Panel (d) shows the normalized charge density $n_p(x,y)-n_e(x,y)$. The supplementary movie 2 animates these figures in time for $-100 \le x \le 0$ and for $15 \le t \le 513$.}\label{plot500} 
\end{figure*}

Figure \ref{plot500} examines in more detail the plasma distribution at the time $t=400$ after the shock front reformed in Fig. \ref{magpressure}. 
Figure \ref{plot500}(a) shows that two-stream modes are present in the interval $-120 \le x \le -30$. They are quasi-planar and their wave vector points on average along $x$. We can estimate their impact on the inflowing upstream particles as follows. Equation \ref{eq:kmax_TS} estimates their wave length as 1.7. A particle with the speed $v\approx 0.25$ is exposed to the electric field of a half-oscillation during 3 time units, where we neglect that the wave is oscillating. The electric acceleration along x is expressed as $dv/dt=E_x$ in normalized units and a particle would change its speed by up to 0.3 during the crossing. The electric field oscillations are accompanied by weak magnetic field oscillations in the same interval in Fig. \ref{plot500}(b). The Weibel instability is responsible for the strong magnetowaves in the interval $-40 \le x \le -20$. Their peak amplitude 0.2 implies that the gyroradius of a particle with the speed 0.25 is of the order unity. The magnetic barrier is thick enough to stop any incoming particle. The oscillations of the magnetic structure in space and time imply that the inflowing upstream particles are scattered into almost random directions. The mean speed of the upstream particles is decreased by the scattering and Fig. \ref{plot500}(c) demonstrates that the slow-down leads to the expected compression. We observe a dense electron beam at $y\approx 4$ and $-55\le x \le -35$. Such beams are usually the product of the filamentation instability but this beam is not driving a magnetic field in Fig. \ref{plot500}(b). A magnetic field is growing in this interval, but the electron beam is not separating patches with a different sign of $B_z$. Figure \ref{plot500}(d) shows that the electron beam is not associated with a net charge; a positron beam with a similar shape cancels the electron charge and current. We had observed a similar electron beam already in Fig. \ref{time55}(c).  

The velocity distribution of the particles can shed further light on the instabilities, which are involved in the thermalization of the incoming upstream flow, and on their impact on the plasma particles. Figure \ref{plot500} reveals 4 distinct plasma regions. The interval $-20 \le x \le 0$ shows only weak wave activity and a high plasma density and it is thus the downstream region of the shock.  The Weibel instability is operational in the interval $-40 \le x \le -20$. The electron density distribution is relatively smooth in the interval $-60 \le x \le -40$ and strong waves are observed for $-80 \le x \le -60$. We integrate the phase space density distribution of the electrons over all y and over all x in the respective interval, which gives us the distribution $f_e(p_x,p_y)$. We also integrate the phase space density distribution over all x of an interval and over all $p_y$, which gives us $f_e(y,p_x)$. 

\begin{figure}
\includegraphics[width=\columnwidth]{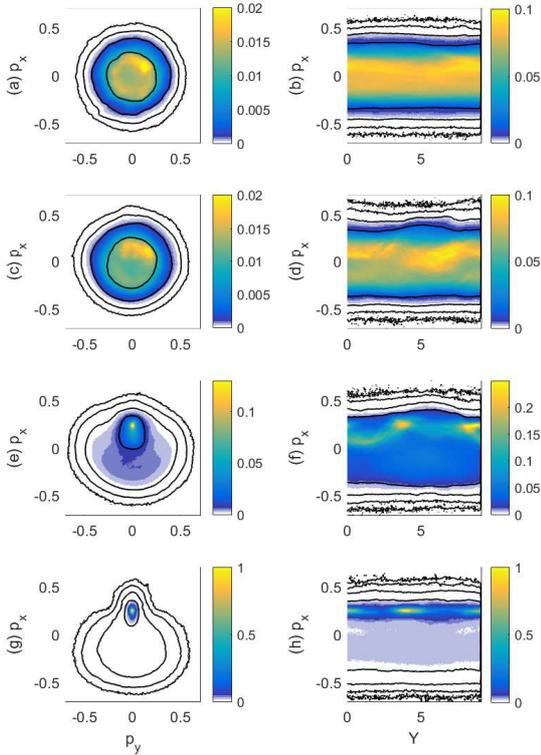}
\caption{The momentum distributions $f_e(p_x,p_y)$ (left column) and $f_e(y,p_x)$ (right column) at the time $t=400$: the distributions in panels (a, b) have been integrated over $-20< x \le 0$. Panels (c, d) show the distributions that have been integrated over $-40<x\le -20$. The distributions in the panels (e, f) have been integrated over $-60<x\le -40$. The distributions, which have been integrated over $-80 < x \le -60$, are shown in panels (g, h). The contours $10^{-5},10^{-4},10^{-3}$ and $10^{-2}$ are overplotted. Each distribution has been normalized first to its total number of particles to remove the bias caused by the different densities in each region. These distributions $f_e(p_x,p_y)$ are then normalized to the peak value in (g), while the distributions $f_e(y,p_x)$ are normalized to the peak value in (h).}\label{velocity500} 
\end{figure}

Figure \ref{velocity500} displays these distributions. 
The electron distribution in the downstream region is practically isotropic in Fig. \ref{velocity500}(a). The density shows a local minimum at $p_x=0$ and $p_y=0$, but the density maximum at ${(p_x^2+p_y^2)}^{0.5} \approx 0.15$ is practically gyrotropic. This distribution is stable against the Weibel instability. The spatial distribution in Fig. \ref{velocity500}(b) is almost uniform. Both distributions match that we expect from a downstream plasma. Figure \ref{velocity500}(c) shows an almost gyrotropic distribution that is superposed on a beam with the mean speed $p_x \approx 0.15$. The thermal energy of this distribution is larger along x than along y and it is unstable to Weibel modes with a wave vector that is aligned with y. The distribution in Fig. \ref{velocity500}(d) oscillates with y. The velocity oscillations are tied to the magnetic pressure gradient force, which is strong here because of the large magnetic pressure $P_B\approx 0.3$. The distribution $f_e(p_x,p_y)$ in Fig. \ref{velocity500}(e) in the interval $-60 \le x \le -40$ is gyrotropic only for the dilute energetic particles. The bulk distribution shows a broad beam centered at $p_x = 0.25$ and $p_y = 0$ and a dilute population that moves in the opposite direction. The latter correspond to the electrons that are emanated by the shock transition layer into the upstream direction. Figure \ref{velocity500}(f) shows a broad and almost uniform particle distribution in the velocity band $-0.3 \le p_x \le 0.3$. A beam is superposed on the distribution in the velocity band $0 \le p_x \le 0.3$. Its mean momentum oscillates along y and we observe localized density maxima at $p_x\approx 0.25$ and y=4 and 9. The particles in this dense beam are those that form the high density striations in Fig. \ref{plot500}(c). The oscillations of the mean momentum along $p_x$ have the same periodicity along y as the localized density accumulations. This connection is clearly established  by the distribution in Fig. \ref{velocity500}(h). The dense band corresponds to the beam of inflowing upstream electrons. These electrons interact with the electrons that escaped from the shock. The velocity distribution in Fig. \ref{velocity500}(g) shows that both electron populations start to form distinct beams. Both beams will separate farther upstream and this distribution is unstable to the filamentation instability. The beamed distribution of the incoming upstream electrons is thus the consequence of a filamentation instability far upstream of the shock. This instability pinches electrons and positrons alike and it does therefore not drive strong magnetic fields upstream (See Fig. \ref{plot500}(b)). The period of the oscillations in Fig. \ref{velocity500}(h) is of the order 4.5, which exceeds by far the lateral width of the simulation box used by \citet{Dieckmann17}. The latter simulation box was thus not large enough to resolve it. The beam diameter is below 1 and thanks to our fine grid its dynamics is resolved well.

\section{Summary}

We have examined the formation of a shock out of the collision of two cold pair clouds at the speed c/2. This speed is close to the maximum one for which the two-stream instability outgrows the filamentation instability. We have demonstrated with the data from the 2D simulation and through its comparison with the results from a 1D simulation that the two-stream instability, which is the only one that grows in the 1D simulation, is responsible for the initial shock formation. The electric fields, which triggered the shock formation, grew in both simulations at the same time. The phase space holes in the 1D simulation were stabilized geometrically, which delayed the formation of the full shock and extended its transition layer. Phase space holes collapse faster if more than one spatial dimension is resolved \citep{Morse969} like in our 2D simulation. The wave number spectrum obtained from the 2D simulation at early times provided further support for the involvement of the two-stream instability in the shock formation.

The electrostatic nature of the two-stream mode required us to resolve with our grid the Debye length of the plasma, which limited the spatio-temporal scale accessible to our simulation. The simulation box was nevertheless large enough to provide insight into the structure of the shock.

We observed the simultaneous growth of three key instabilities that can form in unmagnetized pair plasma, namely the two-stream instability, the filamentation instability and the Weibel instability, albeit in different locations. The two-stream instability developed in a broad layer, which we called the shock transition layer. Their quasi-planarity implied that these quasi-electrostatic waves thermalized the plasma predominantly along the shock normal although their weak magnetic field component would also scatter particles in the perpendicular direction. The pair density and temperature increased slowly over spatial scales of the order of 10 electron skin depths. The thermalization progressed slow enough to facilitate the growth of the Weibel instability in its original form. This instability does not require the presence of multiple particle beams that are separated in velocity space. It grows also in a pair plasma, for which the temperature is higher in one direction than in the others. The Weibel instability resulted in the growth of strong magnetic field patches. These patches were thick enough to block the inflowing upstream plasma and their irregular shape scattered the particles randomly. The plasma density and temperature changed to their respective downstream values over a distance of a few electron skin depths. 

The particle heating in the shock transition layer resulted in a beam of hot pairs that escaped back upstream. Their high temperature results in a decrease of the growth rate of the two-stream instability compared to the filamentation instability and the latter could grow far upstream of the shock. Its effect was to pinch the inflowing plasma orthogonally to the shock normal. The upstream electrons and positrons were compressed into beams with a diameter that was well below the electron skin depth. The matching electron and positron distributions upstream implied that their current was cancelled out to almost zero and these filaments would not yield the strong magnetic field that is observed in simulations of relativistic pair shocks. 

Our simulation confirmed that the filamentation instability is not the one that mediates weakly relativistic shocks. Such shocks are sustained by a combination of the two-stream instability and the Weibel instability in the original form proposed by \citet{Weibel59} and a comparison of our results with those in the simulation by \citet{Dieckmann17} shows that the Weibel instability becomes more important for a higher collision speed.

\section{Acknowledgements}
This work was supported by grants ENE2013-45661-C2-1-P and ENE2016-75703-R from the Ministerio de Educaci\'{o}n y Ciencia, Spain and grant PEII-2014-008-P from the Junta de Comunidades de Castilla-La Mancha. The simulations were performed on resources provided by the Swedish National Infrastructure for Computing (SNIC)  at HPC2N and on resources provided by the Grand Equipement National de Calcul Intensif (GENCI) through grant x2016046960.

\bibliographystyle{mnras}
\bibliography{Manuscript}
\label{lastpage}

\end{document}